\documentclass{article}

\usepackage{arxiv}

\usepackage[utf8]{inputenc} 
\usepackage[T1]{fontenc}    
\usepackage{hyperref}       
\usepackage{url}            
\usepackage{booktabs}       
\usepackage{amsfonts}       
\usepackage{nicefrac}       
\usepackage{microtype}      
\usepackage{lipsum}		
\usepackage{graphicx}
\usepackage{natbib}
\usepackage{doi}

\def\b#1,{{\bf #1,}}

\def\nrdc#1.{1986, in ``Light on Dark Matter'', ed. F.P. Israel,
  (Reidel, Dordrecht), p #1.}
\def\calc#1.{1986, in ``Late Stages of Stellar Evolution'', eds. S. Kwok and
  S.R. Pottasch (Reidel, Dordrecht), p #1.}
\def\camc#1.{1987, in ``Galaxy'', eds. ?. ? (Reidel, Dordrecht), p #1.}
\def\torc#1.{1988, in ``Mass of the Galaxy'', ed. M. Fich,
  (Toronto University Press), p #1.}

\newcommand{\kms}{\,km s$^{-1}$}
\newcommand{\micron}{$\mu\mathrm{}m$}
\newcommand{\arcsec}{$^{\prime\prime}$}
\newcommand{\arcmin}{$^\prime$}
\newcommand{\degr}{$^\circ$}

\title{The remnant and origin of the historical supernova 1181~AD}

\author{ \href{https://orcid.org/0000-0003-0869-4847}{\includegraphics[scale=0.06]{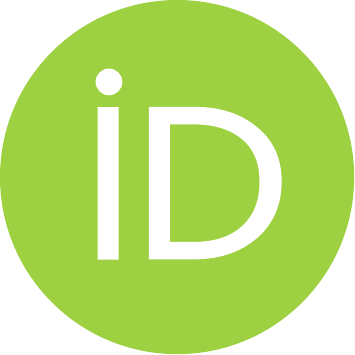}\hspace{1mm}Andreas Ritter} \\
	Department of Physics\\
	The University of Hong Kong,\\
	Chong Yuet Ming Physics Building\\
	Pokfulam Road, Hong Kong\\
	\texttt{aritter@hku.hk} \\
	\And
	\href{https://orcid.org/0000-0002-2062-0173}{\includegraphics[scale=0.06]{orcid.pdf}\hspace{1mm}Quentin A.~Parker}\thanks{Corresponding author} \\
	Department of Physics\\
	The University of Hong Kong,\\
	Chong Yuet Ming Physics Building\\
	Pokfulam Road, Hong Kong\\
	\texttt{quentinp@hku.hk} \\
	\AND
	\href{https://orcid.org/0000-0002-6394-8013}{\includegraphics[scale=0.06]{orcid.pdf}\hspace{1mm}Foteini~Lykou}\\
    Konkoly Observatory\\
    Research Centre for Astronomy\\
    and Earth Sciences\\
    Konkoly-Thege Mikl\'os \'ut 15-17\\
    1121 Budapest, Hungary\\
	\texttt{foteini.lykou@csfk.org} \\
	\AND
	\href{https://orcid.org/0000-0002-3171-5469}{\includegraphics[scale=0.06]{orcid.pdf}\hspace{1mm}Albert A.~Zijlstra}\\
	Jodrell Bank Centre for Astrophysics\\
	The University of Manchester\\
	Oxford Road, M13 9PL\\
	Manchester, UK\\
	\texttt{albert.zijlstra@manchester.ac.uk} \\
	\AND
	\href{https://orcid.org/0000-0002-7759-106X}{\includegraphics[scale=0.06]{orcid.pdf}\hspace{1mm}Mart\'in A.~Guerrero}\\
	Instituto de Astrof\'isica\\
	de Andaluc\'ia (IAA-CSIC)\\
	Glorieta de la Astronom\'ia S/N\\
	18008 Granada, Spain\\
	\texttt{mar@iaa.es} \\
	\AND
	Pascal~Le D\^u\\
	Kermerrien Observatory\\
	F-29840 Porspoder, France\\
	\texttt{pascal.le.du@shom.fr}
}

\renewcommand{\shorttitle}{SN 1181 AD}

\hypersetup{
pdftitle={The remnant and origin of the historical supernova 1181~AD},
pdfsubject={supernova remnants},
pdfauthor={Andreas Ritter, Quentin A. Parker, Foteini Lykou, Albert A. Zijlstra, Mart\'in A.~Guerrero, Pascal~Le D\^u},
pdfkeywords={(stars:) supernovae: general, (stars:) supernovae: individual: historical, stars: Wolf–Rayet, ISM: supernova remnants},
}

\begin{document}
\maketitle

\begin{abstract}
	The guest star of AD 1181 is the only historical supernova of the last millennium that is without a definite counterpart. The previously proposed association with supernova remnant G130.7+3.1 (3C58) is in strong doubt because of the inferred age of this remnant. Here we report a new identification of SN~1181 with our co-discovery of the hottest known Wolf-Rayet star of the Oxygen sequence (dubbed “Parker’s star”) and its surrounding nebula Pa\,30. Our spectroscopy of the nebula shows a fast shock with extreme velocities of $\approx$1,100\kms. The derived expansion age of the nebula implies an explosive event $\approx$1,000 years ago which agrees with the 1181 event. The on-sky location also fits the historical Chinese and Japanese reports of SN~1181 to 3.5~degrees. Pa\,30 and Parker's star have previously been proposed to be the result of a double-degenerate merger, leading to a rare Type~Iax supernova. The likely historical magnitude and the distance suggest the event was subluminous for normal supernova. This agrees with the proposed Type~Iax association which would also be the first of its kind in the Galaxy. Taken together, the age, location, event magnitude and duration elevate Pa\,30 to prime position as the counterpart of SN 1181. This source is the only Type~Iax supernova where detailed studies of the remnant star and nebula are possible. It provides strong observational support for the double-degenerate merger scenario for Type~Iax supernovae.
\end{abstract}

\keywords{(stars:) supernovae: general \and (stars:) supernovae: individual: historical \and stars: Wolf–Rayet \and ISM: supernova remnants}

\section{Introduction}
Only 9 historically recorded supernova (SN) explosions are known in the Galaxy \citep{green2002}. In only 5 cases has the remnant of the supernova been identified. For the other cases, the remnant is not known with certainty.  The remnant is crucial for identifying the type of supernova, whilst the known time of the explosion and duration constrain the models of the evolution of the remnant.

Here, we consider the historical `Guest Star' of  1181~AD recorded by Chinese and Japanese astronomers \citep{2020AN....341...79H}; it remained visible for 185 days from Aug 6, 1181 to Feb 6, 1182~AD \citep{Hsi1957}. The supernova remnant (SNR) or pulsar wind nebula G130.7+3.1 (3C58, which we will hereafter use to refer to this SNR) are located in its vicinity \citep{stephenson1999,green2002} and have till now been considered as linked with SN~1181. However, a precise estimate of the expansion age of this nebula of 7,000 years, based on NVSS radio observations over 20~years, and the spin-down age of the pulsar (5,400 years) by \citet{2006ApJ...645.1180B} have put the association in serious doubt, although not completely excluding it \citep{Kothes2013}. A strong argument for this association was that until now there was no other candidate known for the remnant. This left SN~1181 as the youngest Galactic supernova without a firmly confirmed remnant. We propose that the recently discovered nebula Pa\,30 and its extreme central star \citep{gvaramadze2019} are in fact the remnant and residual core of the 1181~AD explosion.

\begin{figure*}[tbph]
	\centering
	\includegraphics[width=\textwidth]{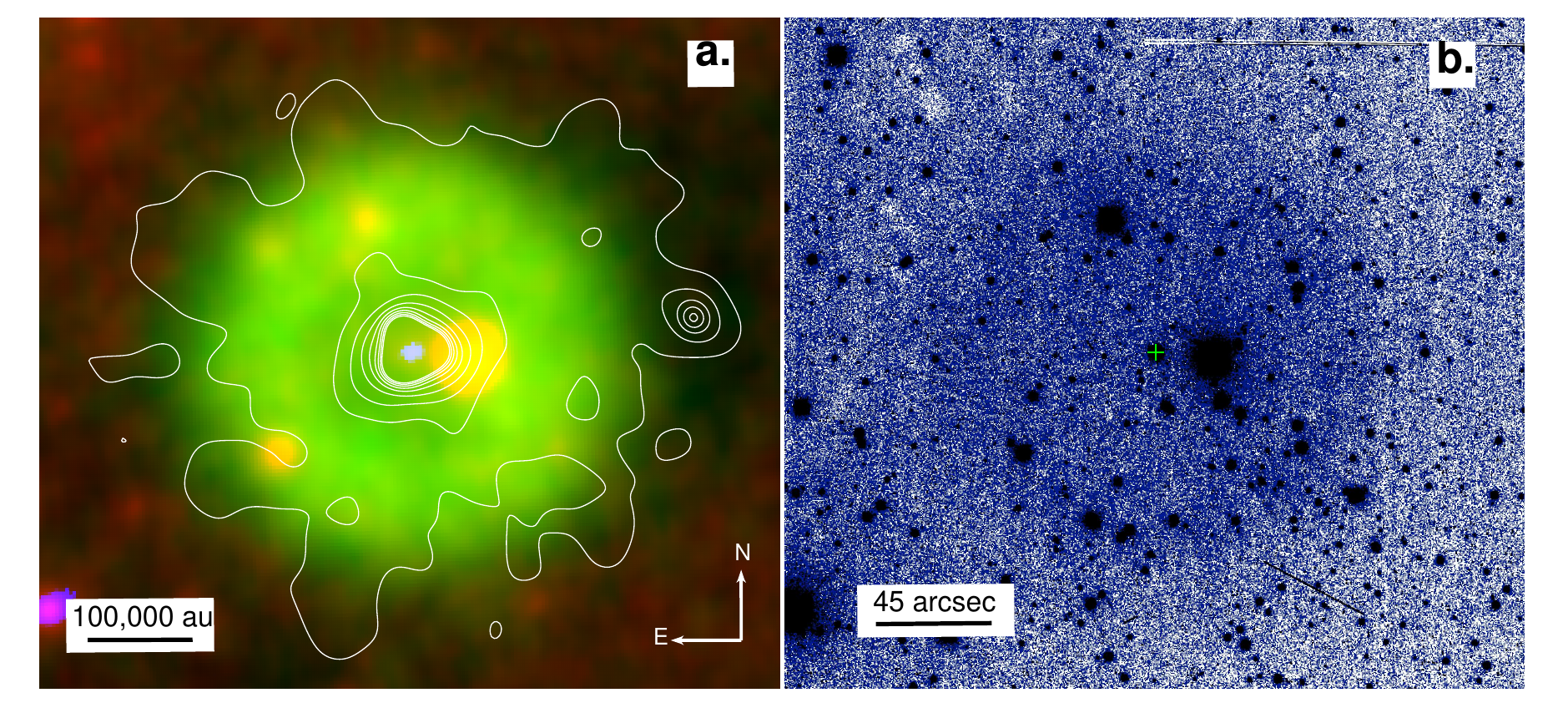}
	\caption{\small {\bf (a.)} False colour image of Pa\,30 where {\it red} stands for \emph{WISE} 11~micron, {\it green} for \emph{WISE} 22~micron, and {\it blue} for \emph{GALEX} near-UV, with \emph{XMM-Newton} contours (10 levels, linear scale). Only the CS is visible in the near-UV and the majority of the X-ray emission originates from the core of the nebula. A background point source is seen westward of the CS in the \emph{XMM-Newton} contour map. {\bf (b.)} 2.1-m KPNO [O~{\sc iii}] image, which we have stacked and rebinned from individual frames to enhance the low surface brightness, diffuse shell. The green cross in the center of the image marks the location of the CS. Panels (a) and (b) are reproduced at the same angular scale. At the \emph{Gaia} distance of Pa\,30 of $2.30\pm0.14$\,kpc, an angular scale of 45\arcsec\ translates to about 100,000~AU.
	}
	\label{fig:colour}
\end{figure*}

The nebula Pa\,30 was discovered on 25$^{\rm th}$ August 2013 from the Wide-field Infrared Survey Explorer (\emph{WISE}) mid-IR image archive \citep{wise} by Dana Patchick \citep{kronberger2016} from our affiliated `Deep Sky Hunters' (DSH) amateur astronomer group, and is included in the HASH PN database\footnote{\url{http://hashpn.space}} \citep{hash} as Pa\,30. The \emph{WISE}  W3 (11\micron) band  shows a  round nebula, whereas the dominant shape in the  W4 (22\micron) band is donut-like inside a much fainter halo. Narrow-band [O~{\sc iii}] imaging observations obtained by the DSH group on the 2.1m KPNO telescope in September 2013 indicate a very faint, diffuse, circular emission feature. 
The nebula is a strong source of diffuse X-ray emission \citep{oskinova2020}. 
Multi-wavelength images of the nebula are shown in Figure~\ref{fig:colour}.   The bright central star (CS) is both hydrogen-poor and helium-poor, and has a unique emission-line spectrum  as shown by \cite{gvaramadze2019} and from our independent spectroscopy.

\begin{figure*}
	\centering
		\includegraphics[scale=0.95]{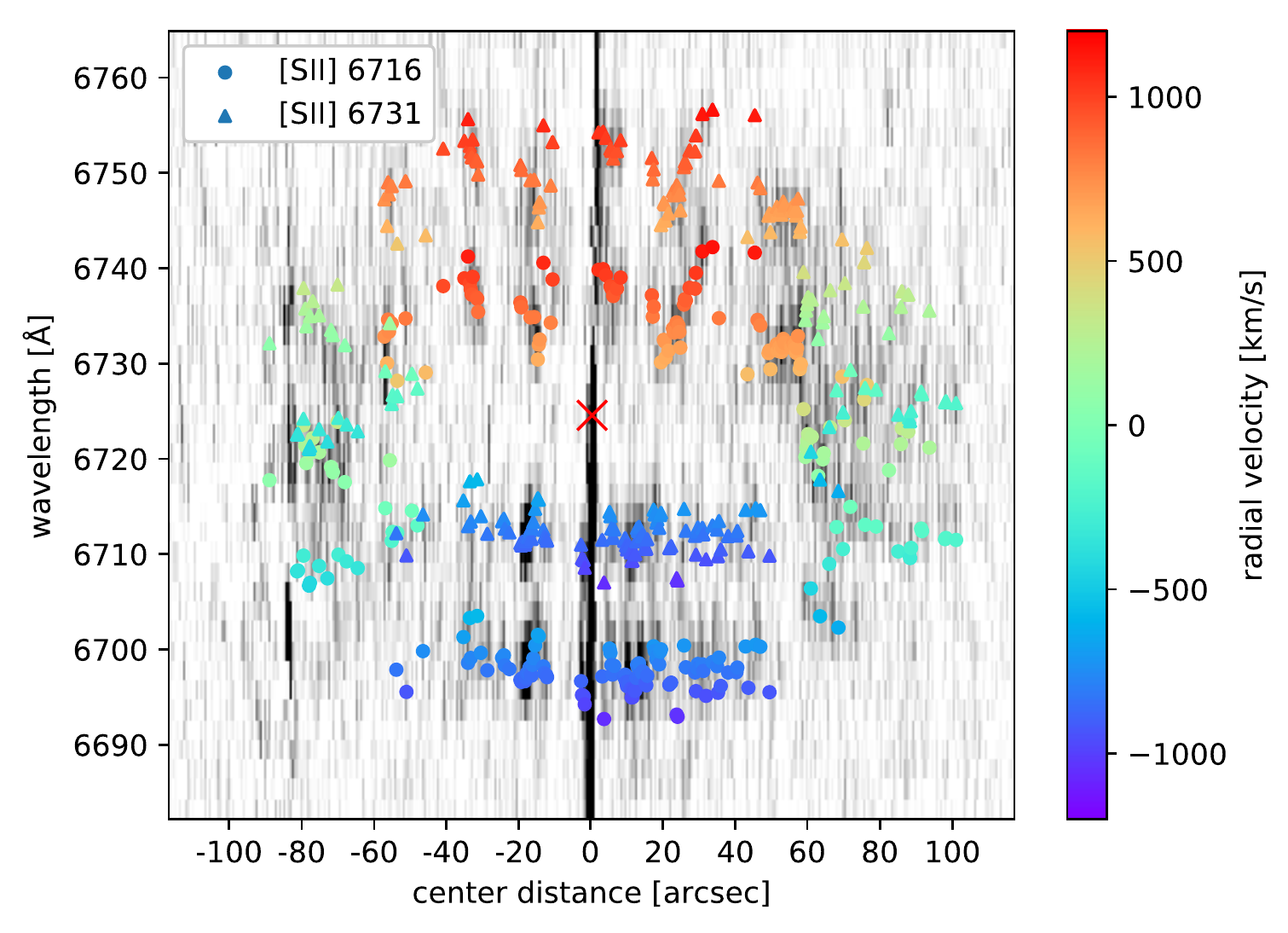}
	\caption{\small
GTC/OSIRIS position-velocity diagram of the [S {\sc ii}] lines (circles: 6716\AA, triangles: 6731\AA; grey scale in arbitrary data units) along the slit, overplotted with the measured line positions, and color-coded to the radial velocity for each measured line. Left is NNE of the CS, right is SSW. The position of the CS is marked with a red `X'. Spectral features have been enhanced by subtracting the mean of each spectrum above and below the [S {\sc ii}] lines. 
The image shows sharp velocity variations with increasing distance from the CS at $\approx-$20\arcsec, $-$40\arcsec\ and +25\arcsec\, where the [S {\sc ii}] lines are also much brighter than for the rest of the long-slit spectrum, indicating strong shock excitation both for the near as well as the far side of the nebula. 
}
	\label{fig:vrad}
\end{figure*}

On careful scrutiny both sets of our observations (see below) revealed two faint optical emission lines species in the nebula: the [S {\sc ii}] 6716 and 6731\AA\ doublet, and the [Ar {\sc iii}] 7136\AA\ line. The  [S {\sc ii}] in particular shows extreme velocity splitting with an expansion velocity up to $v_{\rm exp}\approx$ 1,100 $\pm$ 100 \kms\ (Fig.~\ref{fig:vrad}). The velocity is constant up to 50\arcsec\ from the CS, followed by a sharp decline to systemic velocity at $\approx$100\arcsec. We also extracted the star's spectrum after it was first brought to our attention by Pascal Le D\^u, an amateur collaborator of the corresponding author, where its extreme nature became clear \citep[e.g. see][]{gvaramadze2019} and it was thereafter dubbed ``Parker's" star.

Taking the latest {\it Gaia} distance of $2.30\pm0.14$\,kpc \citep{bailerjones2021}, the highest velocity extends to a total diameter of $2.2 \pm0.4$ pc, and the slower emission extends to about twice that diameter. If we assume that the 1,100\kms\ gas is expanding at a constant velocity, and that the uncertainty on the expansion velocity is 10\%, the highest velocity shell has a kinematic age of $990^{+280}_{-220}$ years. 

\citet{oskinova2020} estimated an age for the shell based on the Sedov solution scaling relation \citep[see][]{borkowski2001, oskinova2005} for supernova remnants of $\sim$1,000 years, in excellent agreement with the measured kinematic age of the shell.  This agreement supports the identification of the nebula as an SNR. 

The kinematic age agrees well with SN~1181 recorded 840 years ago and provides a strong temporal argument for association. The second argument for the association comes from the position concordance. Of the recorded `guest stars' listed in \cite{2020AN....341...79H}, SN~1181 is the only one that matches the estimated explosion date and the location of Pa\,30 to within the errors. Averaging the five reported positions for this historical SN (\citeauthor{2020AN....341...79H}) and transforming the average to J2000, the separation on the sky between the recorded position and the position of Parker's star is only 3.5 degrees (see Table~1), well within the uncertainties. The {\it Gaia} proper motion for the star of 2.7~milliarcsec\,yr$^{-1}$ would only shift it by $\sim$3~arcseconds in the intervening period. The only other viable alternative \citep[and indeed the previously favoured association,][]{Kothes2013} is SNR 3C58 at $\sim$4.5~degrees away from the SN~1181 position.

\begin{figure*}[htbp]
	\centering
	\includegraphics[width=\textwidth]{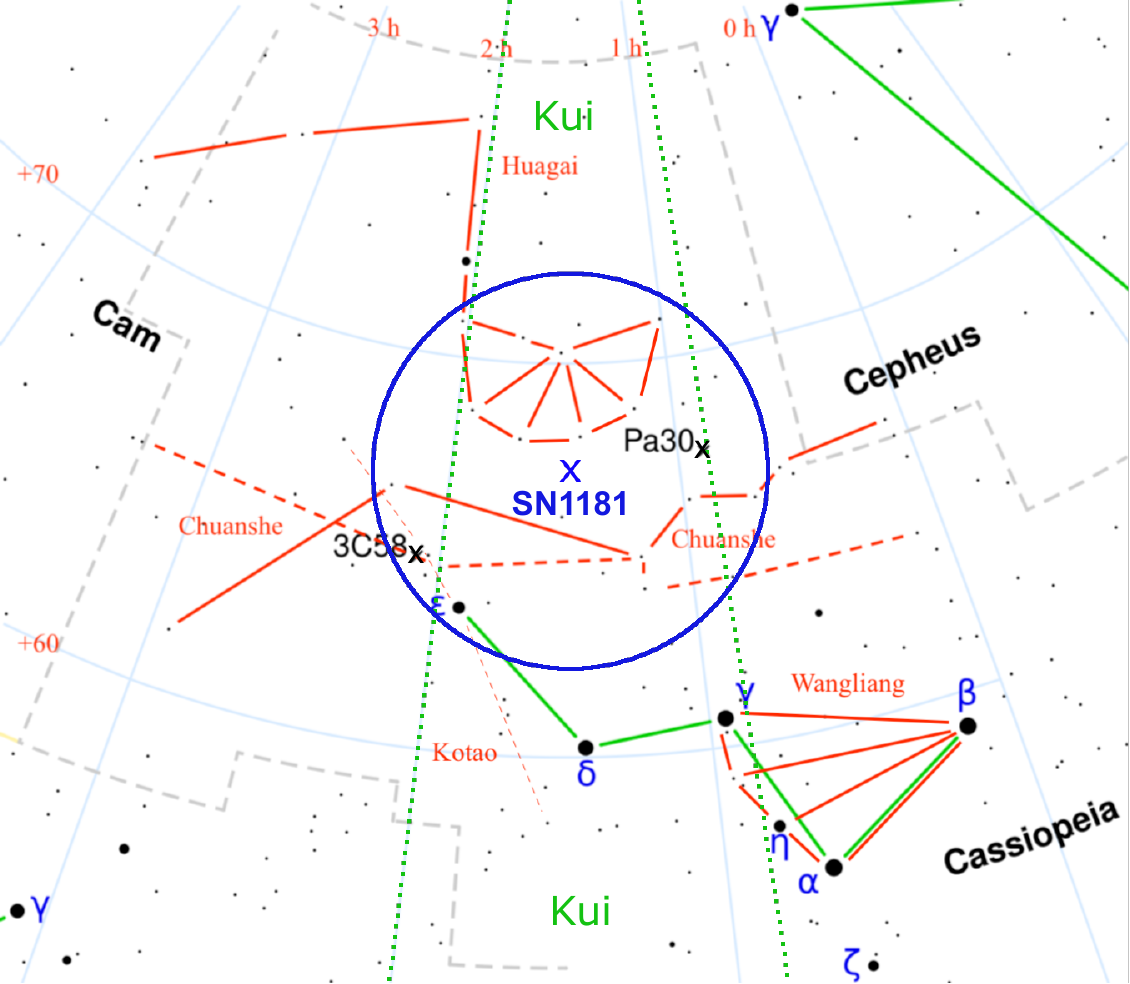}
	\caption{\small The region of SN~1181, as described in \cite{Hsi1957} with the asterisms according to \citet[their Fig A5]{2020AN....341...79H} and \citet[their Fig 9.1. Red dashed line shows their asterism for Chuanshe]{Stephenson2002} indicated.  Pa\,30 and 3C58 are indicated by black crosses.  The green line indicates Cassiopeia. The Chinese lunar lodge Kui is between the two green dotted lines. The supernova was stated to lie between Huagai and Chuanshe, near Wangliang. The best estimated average position of SN~1181 is given by a blue cross surrounded by a blue error circle of radius 5~degrees. The background star chart was created using \href{pp3.sourceforge.net}{PP3 - Celestial Chart Generation}.
	}
	\label{fig:map}
\end{figure*}

Figure \ref{fig:map} shows the best estimated general location of SN~1181. The two candidates Pa\,30 and 3C58 are indicated, as is the best estimate for the location of SN~1181 with a 5~degree radius error circle.  The SN was reported to have occurred near Wangliang in Huagai, invading Chuanshe \citep[spelling adopted from \citeauthor{2020AN....341...79H}]{Hsi1957}. These Chinese asterisms  are shown by the three red-lined constellations, as taken from \citeauthor{2020AN....341...79H} The location of Chuanshe is not known with full certainty: the dashed line shows the location according to \cite{Stephenson2002}. The text suggests that SN~1181 should be between the two constellations.  Pa\,30  fits this well. The 3C58 alternative is located in or south of Chuanshe, not obviously associated with Huagai. We also note that the location of 3C58 would have been more easily described as close to the bright (3rd magnitude) star $\varepsilon$ Cas.

The magnitude of SN~1181 is not known with certainty but it was compared in the manuscripts to Saturn. With the reasonable assumption that this refers to its brightness, then the magnitude at peak may have been around $m_V\sim -0.5$ to $+1.0$. This agrees with the fact it was obvious enough to be picked up in China and Japan who at the time had well developed astronomical capacity, but was missed in others (e.g. Korea and Europe). For a distance of 2,300 pc and an extinction $A_V=2.4$ (see Methods), the absolute magnitude becomes $M_V \sim -14$ to $-12.5$. This is significantly sub-luminous for a typical supernova but is within the range for Type~Iax events (see below). The object also remained visible for 185 days. Given that the limit of naked eye visibility is $m\approx 5.5$, this indicates 4.5 -- 6 magnitudes of fading over this period. The long period of naked eye visibility is typical for a supernova but  does not fit a typical nova duration of days to a few weeks.

Pa\,30 is a source of diffuse X-ray emission with a spatial extent in serendipitous \emph{Swift} XRT and pointed \emph{XMM-Newton} EPIC observations larger than that in the mid-IR \emph{WISE} and optical [O~{\sc iii}] images (Fig.~\ref{fig:colour}). 
The \emph{XMM-Newton} EPIC X-ray spectra of both the CS and nebula have been fitted with plasma emission models with significant enrichment in neon, magnesium, silicon, and sulphur \citep{oskinova2020}, which has been interpreted as the result of the incomplete carbon and oxygen fusion expected in Type~Ia SNe. 

Pa\,30 can accordingly be classified as a supernova remnant \citep{gvaramadze2019,oskinova2020}, but the bright CS, which lacks hydrogen and helium, shows it is not a common type. Instead it points at a Type~Iax supernova as suggested by \citeauthor{oskinova2020}, that is, a sub-luminous SNe Ia event in which the star does not self-destruct. While the exact mechanisms leading to a Type~Iax SN are still not fully understood, they are believed to arise from either the failed detonation of a Carbon-Oxygen (CO) White Dwarf (WD) accreting material from a Helium donor star \citep[single degenerate scenario, e.g. ][]{kromer2015,jordan2012} or from a CO WD merging with a heavier Oxygen-Neon (ONe) WD \citep[double degenerate scenario, ][]{kashyap2018} where the accretion disk itself deflagrates. 

A key discriminant between typical SN of Type~Ia and Type~Iax is the expansion velocity with $\sim$10,000\kms\ for Type~Ia and only 2,000--7,000\kms\ for Type~Iax \citep{jha2017}. Our measured expansion velocity for the nebula is 1,100\kms, which argues in favour of a Type~Iax event. Type~Iax show  a large range of peak magnitudes, which at the faint end ($-13$ to $-14$) \citep{jha2017} overlap with the likely magnitude for SN~1181 at a distance of 2,300 pc. This also independently supports this classification.

Based on the current stellar luminosity, Pa\,30 is likely a  double degenerate merger (\citeauthor{gvaramadze2019}) where \citeauthor{oskinova2020} interpret it as a ONe-CO WD merger based on the nebula's neon abundance. Models for high-mass ONe-CO mergers \citep{kashyap2018} predict a faint SN~Iax with an absolute magnitude of only $-11.3$ mag. This is a bit fainter than what we find for SN~1181 but indicates that the \citeauthor{oskinova2020} interpretation may be supported by the faint magnitude of the event.

\section{Methods}

\subsection{Coordinates}

The likely position of SN~1181 based on the historical descriptions is reported by \cite{2020AN....341...79H} as  RA: 01:31:14  DEC: +70:15:50 (J2000) with an uncertainty of $\sim$4 degrees (their Table 5). They also report previous determinations by four authors.
We averaged all five positions, converted to J2000 coordinates. This gives RA: 01:29:04 DEC: +67:15:33. The uncertainty on the position is taken as 5 degrees.

\begin{table}
	\centering
	\caption{Coordinates (J2000). The original SN~1181 coordinates (e.g. 1:30, +65) have been precessed from B1950 to J2000. The uncertainty in these values remains high. The last two columns show the distance to the two candidate remnants in degrees. References to the historical positions are listed in \cite{2020AN....341...79H} .}
	\label{tab:coord}
	\begin{tabular}{cllcc}
           & RA & DEC &  $\Delta$(Pa\,30) & $\Delta$(3C58) \\ \hline
          \multicolumn{5}{c}{\it SN 1181 coordinates} \\
          Stephenson & 01:33:30 & +65:15:23 & 4.6 & 3.4\\
          Hsi  & 01:43:53 &+70:15:05   & 5.3 & 5.8 \\
          Psovskii & 01:03:12 & +65:16:03 & 2.4 & 6.6  \\
          Xi & 01:33:30 & +65:15:23   & 4.6 & 3.4 \\
          Hoffmann & 01:31:14 & +70:15:53 & 4.4 & 6.3 \\
          {\it average} & 01:29:04 & +67:15:33 & 3.5 & 4.4 \\
          \multicolumn{5}{c}{\it Candidate remnant coordinates}\\
          Pa\,30 & 00:53:11.2 & +67:30:02.4 \\
          3C58 & 02:05:37 & +64:49:42 \\
          \hline
	\end{tabular}
\end{table}

The proposed coordinates of SN~1181 (accurate to 5 degrees) and the two candidate remnants are listed in  Table \ref{tab:coord}. Based on the offsets from the proposed coordinates, both candidates are within the positional uncertainty, and on this criterion alone each appears viable. 

Following \cite{Hsi1957} and \citet{2020AN....341...79H}, we have compared the locations of both candidates to the original Chinese and Japanese reports described by \citet{Stephenson2002}. Two texts from China place the guest star `in Kui lunar lodge, trespassing against Chuanshe and guarding the 5th star of Chuanshe' (South China) and `in Huagai' (North China). A record from Japan describes it `close to Wangliang and guarding Chuanshe'. The Kui lunar lodge and the three asterisms are depicted in Fig.~\ref{fig:map}, which is based on Fig.~A5 of \citet[][we note that the declination scale in their Fig.~A5 is incorrect]{2020AN....341...79H}. Wangliang contains the brightest stars in the area being part of the modern constellation Cassiopeia. The seven stars of Huagai, although much fainter, form a well-defined cluster shaped like a parasol and are also fairly easily identified. The exact position of Chuanshe is less certain. This asterism consists of nine dim stars barely visible to the unaided eye. Two possible positions (\citeauthor{2020AN....341...79H}, their Fig.~A5, and \citeauthor{Stephenson2002}, their Fig.~9.1) are shown in Fig.~\ref{fig:map}. Taken together, the supernova can be expected to have occurred in between Huagai and Chuanshe. Our averaged best estimate of its position is given by a blue cross centred on a blue error circle of radius 5 degrees. It is notable that none of the records mention the asterism Kotao (thin dashed line in Fig.~\ref{fig:map}), although it crosses Chuanshe at the location of 3C58. One would also expect that the location of 3C58 would have been better described as being close to $\varepsilon$ Cas, a bright 3rd magnitude star. In contrast, the location of Pa\,30, also denoted by a labelled black cross, fits the description well and is in better agreement with its association with Huagai. However, the descriptions are not accurate enough to decide between the two candidates on positional information alone though Pa\,30 is clearly favoured.

\subsection{Observations}
We obtained optical spectroscopy of both the star and the nebula on 2016 July 8 using long slit observations with the OSIRIS instrument of the 10-m GranTeCan (GTC) telescope \cite{Cepa2010} using the grism R1000B. The 7.4\arcmin\ slit with a width of 0.8\arcsec\ was placed at a position angle 30\degr\ East-of-North, slightly off-centre from the central star to reduce contamination from a bright field star. Total exposure time was $2 \times 20$ min. A factor $2\times2$ binning was used which provides a spatial scale of 0.254\arcsec/pixel (with a seeing of 1.35\arcsec) and a spectral resolution of about 2\AA\ (R=1,000). The wavelength coverage was 3700--7000\AA. We used the standard GTCMOS pipeline for the data reduction.

Earlier, the DSH team used the SparsePak integral field unit (IFU) of the Bench mounted spectrograph on the 3.5-m WIYN telescope at KPNO on 2014 October 15 to observe Pa\,30 as a planetary nebula candidate. They did not see the expected emission lines and did not extract the central star's spectrum. The exposure time was $2\times20$min. The IFU mode offers 82 fibers of 4.7\arcsec\ diameter,  in a 72\arcsec$\times$71\arcsec\ grid \citep{sparsepak}. The wavelength coverage with grating 600V (10.1 degree Littrow blaze angle) was 4280--7095\AA\ giving a spectral resolution of 3.35\AA.  We re-reduced this archived data using the pipeline of \cite{ritter2004} and standard {\sc iraf} packages extracting both the star and nebula. Since the dedicated sky fibers were also placed on regions with potential nebular emission, we subtracted the sky using standard principal component analysis. 

\subsection{Interstellar extinction }\label{sec:gaia}

The neutral hydrogen column density estimates \citep{HI4PI} yield an extinction of $A_V = 2.24\pm0.1$ mag. The higher resolution 3-D IPHAS extinction map \citep{sale2014} indicates a higher extinction of $A_0=2.52\pm0.40$ mag (monochromatic). 

We investigated 3-D extinction maps based on the {\it Gaia}-2MASS \citep{lallement2019} and {\it Gaia}-PanSTARRS-2MASS \citep{green2019} databases. In the first case (\citeauthor{lallement2019}), we find a colour excess $E(B-V)=0.87$ and therefore $A_V=2.7$ mag. In the latter case (\citeauthor{green2019}), $E(g-r)=0.70$ was found at 2.3 kpc. $E(B-V)/E(g-r) =0.884$ to 0.996, therefore $A_V$ ranges between 1.92 and 2.17 mag.

We also selected stars within 10~arcminutes of Pa~30 in the {\it Gaia} EDR3 catalogue with parallax uncertainty $\sigma_\pi /\pi < 0.5$ with a non-zero $A_G$ from {\it Gaia} DR2. The median $A_G$ is about 2.25 mag. Using Table 13 of \cite{jordi2010}, we extrapolate $A_G / A_V$ ratios between 0.88 and 1.02 for stars hotter than 50kK, giving an interstellar extinction of $A_V=2.4\pm 0.2$ which we have adopted for dereddening.


\section{Conclusions}

The previous association of  3C58 with SN~1181 was based in part on the lack of another viable candidate, but it has a difficulty with the discrepant ages from the observed expansion velocity, proper motion of the knots, neutron star cooling models, and pulsar spin down rate \citep{2006ApJ...645.1180B, fesen2008}. \cite{Kothes2013} proposed a much smaller distance to 3C58 than previously found (2~kpc versus 10~kpc), which reduces the age discrepancy from the expansion velocity, but this would then imply an event with a peak apparent magnitude of $m=-5$. Such a bright event disagrees with the reported comparison with Saturn, and leaves the question open why it wasn't seen in more countries. Pa\,30 now provides an excellent, viable candidate for the SN~1181 eruption that fits the location, the age, the brightness and even the visible duration given its likely Type~Iax nature.

We therefore conclude that Pa\,30 is the remnant of the SN~1181 supernova. It was until now the only remaining historical supernovae of the last millennium without a certain counterpart. It is also the first observed supernova Type~Iax event in the Galaxy and so the only one where detailed studies of the remnant star and nebula are possible, and for which the double-degenerate merger scenario has strong observational support \citep{gvaramadze2019,oskinova2020}. Given the extreme nature of Parker's star itself \citep{gvaramadze2019,oskinova2020}  and our linking it to the 1181~AD supernova, this source is of considerable scientific and historical interest. 
Parker's star is the only Oxygen Wolf Rayet star known that is neither the result of a massive Pop~I progenitor nor the central star of a planetary nebula, but is the result of an ONe-CO WD merger that accompanied a Type~Iax supernova explosion that now has an historical basis.


\section*{Acknowledgements}
We acknowledge with thanks Matthias Kronberger and Dana Patchick of the Deep Sky Hunters amateur group for their dedication in discovering new PNe and uncovering Pa\,30. We also thank Dr. Laurence Sabin (UNAM) for designing the GTC/OSIRIS observations on behalf of other team members Q.A.P and A.A.Z.  Q.A.P thanks the Hong Kong Research Grants Council for GRF research support under grants 17326116 and 17300417. A.R. and F.L. thank HKU for the provision of  postdoctoral fellowships, while FL also acknowledges funding from MTA Hungary (OTKA project no. 132406). A.A.Z. thanks the Hung Hing Ying Foundation for the provision of a visiting professorship at HKU and acknowledges funding from the UK STFC under grant ST/T000414/1. M.A.G. was funded under grant number PGC2018-102184-B-I00 of the Ministerio de Educaci\'on, Innovaci\'on y Universidades cofunded with FEDER funds. 

This work made use of the extinction map query page, hosted by the Centre for Astrophysics and Planetary Science at the University of Kent; and of data products from the Wide-field Infrared Survey Explorer, which is a joint project of the University of California, Los Angeles, and the Jet Propulsion Laboratory/California Institute of Technology, funded by the National Aeronautics and Space Administration. Based on observations obtained with \emph{XMM-Newton}, an ESA science mission with instruments and contributions directly funded by ESA Member States and NASA.

%

\vspace{5mm}
\textit{Facilities}: WISE, GALEX(near UV), XMM(Newton), GTC:10.0m,OSIRIS
KPNO:WIYN3.5m,SparsePak, KPNO:2.1m\\

\vspace{5mm}
\textit{Software}: astropy \citep{astropy:2013,astropy:2018},  
          \href{pp3.sourceforge.net}{PP3 - Celestial Chart Generation} created by Torsten Bronger,
          \href{inkscape.org}{Inkscape} that is Free and Open Source Software licensed under the GPL



\bibliographystyle{unsrtnat}
\bibliography{references}  






\end{document}